\documentclass[aps,prl,preprint]{revtex4}
\begin{document}

\title{ The zero mass limit of Kerr and Kerr-(anti) de Sitter space-times: Revisited}

\author{M. Horta\c{c}su \footnote{E-mail:hortacsu@itu.edu.tr}}
\affiliation{Mimar Sinan G\"{u}zel Sanatlar \"{U}niversitesi, Istanbul, Turkey} 

\date{\today}
https://tr.overleaf.com/project/5cf65feff225202797222ebf
\begin{abstract}
We continue studying the zero mass limit of the Kerr-(Anti) de Sitter space-times by
investigating the possibility of
special values of the frequencies to have polynomial solutions for the radial wave equation, and
compute  the reflection  coefficients at the origin 
for waves  coming from the infinity for the AdS and from the cosmological horizon in the dS cases. . 
\end{abstract}

\pacs{04.70.-s, 04.62.+v, 02.30.6p}

\keywords{Kerr metric, wave equation, exact solutions}
\maketitle

\section{Introduction}

In a very interesting paper Gibbons and Volkov
\cite{Gibbons} have claimed that the mass going to zero limit of the Kerr \cite{Kerr}
metric has wormhole solutions. This claim  does not agree with  contrary claims \cite{Landau,Visser}.
In an earlier paper \cite{tolga},
we gave one explicit
solution  for a scalar particle coupled to the zero mass of the
limit of both  the Kerr,  Kerr-dS  and Kerr-AdS space times, using the wave equation given
by Gibbons and Volkov
\cite{Gibbons}.
In that work we had
confirmed the result of Gibbons and Volkov by explicitly obtaining one of
the wave solutions, which has a cut singularity along the z- axis for the angular equation, and at the origin for the radial equation. These
solutions brought to attention whether, in this model, one can calculate the scattering at the origin and   normal modes.

In this paper, we continue studying the properties of the wave equation using the radial variable
$u=-\frac{r^{2}}{a^{2}}$,
for a scalar particle in this metric.  The interesting aspect of the metric used is that with
appropriate choices of the transformation of
the radial and polar angle  coordinates, we get the same wave equation for both coordinates.

Actually, the wave equation has two solutions. One of the solutions validates the Gibbons-Volkov result,
with a singularity at the origin
and along the z- axis, which is interpreted as a wormhole in  \cite {Gibbons}. The other solution is smooth at the origin, and at the z- axis.
As expected,
there is no event horizon for the Kerr case in the zero mass
limit of the black hole.

We have
to report that there was an unfortunate mistake in eq.(1) of the
Correction \cite{tolga1} to \cite{tolga}, where the term without
derivative should be divided by $4$. Fortunately this error does not change the  essential result 
given in the solution, given in eq.(19), of \cite{tolga}.

Here we complete this work by
finding  the possible normal modes to obtain the polynomial solutions for this equation, studying the
possible reflection of
 waves coming from  the cosmological horizon (dS) and from infinity (AdS) towards the origin. In an appendix, we study the solution at infinity, valid for the AdS solution. Since we could not produce a formula for connecting the solution expanded around infinity to the solution expanded around zero, we put this part to  Appendix A. We, however, with a transformation,  bring the point at  infinity to  the point at one, and compute the reflection coefficient. In  Appendix B, we  derive  a connecting formula for Heun functions at zero and at an arbitrary finite point. In appendix C, we give solutions which are convergent at two singular points of the solutions  for the de Sitter  case, by changing the independent variable in the respective  equation. The similar analysis is done in the main text for the AdS case.

\section{2. Solutions and Normal Modes}

As given in \cite{Gibbons}, if we take the time dependence as $e^{i\omega t}$, and  $\phi$ dependence
as $e^{-im\phi}$, our wave
equation for a scalar particle separates into two equations and  read:
\begin{eqnarray}
&&\frac{d}{dr} \bigg((r^{2} + a^{2} ) D \frac{dF(r)}{dr}\bigg)
+ \bigg( \frac{3 \Xi \omega^2}{\Lambda D} -\mu^{2} r^{2}
+\frac{m^{2} a^{2} \Xi}{ r^{2} + a^{2}} + \lambda \bigg) F(r) =0,
\end{eqnarray}
\begin{eqnarray}
&&\frac{1}{\sin\theta} \frac{d}{d \theta}\bigg(\sin{\theta} C \frac{dG (\theta)}{d \theta} \bigg)
+\bigg(\frac{3 \Xi\omega^2}{\Lambda C} + \mu^{2} a^{2} \cos^{2} {\theta}  + \frac{ m^{2} \Xi }{\sin^{2}\theta}
 + \lambda \bigg)G(\theta) = 0.
\end{eqnarray}
Here $ D= 1-{\frac{ \Lambda r^{2} }{3}}$, $C= 1+ {\frac{{ \Lambda a^{2} }}{{3}}\cos^2\theta}$ and
$\Xi= 1 + {\frac{{\Lambda a^2 }}{{3}}}$. The independent variables $\theta$ and its functions, and $r$ appear in the angular and radial equations with even powers.
Thus, any solution which is valid for positive $\theta$ and  $r$ are also valid for their negative values.
If we make the transformations $x= \cos^{2}\theta$ and  $ u= - r^2/a^2 $, we find that  we  get exactly the
same expressions
for both the radial and the angular
equations.  

One may wonder why we define $u$ between zero and minus infinity. We make the  transformation
for $u$ in this way, just to
be able to fit to the standard Heun form \cite{Heun,Ronveaux,Slavyanov,Hortacsu,Tolga} .
\begin{eqnarray}
&&\frac{d^{2}H_G (x)}{d x^{2}} +\bigg( \frac{\delta}{x-1} + \frac{\epsilon}{x-b}+ \frac{{\gamma}}{{x}}
\bigg) {\frac{{dH_G (x) }}{{d x }}}
+\bigg(\frac{{\alpha \beta x-q}}{{x(x-1)(x-b)}} \bigg) H_G(x) = 0.
\end{eqnarray}

Our eq.(1), with the transformation to $u$ reads
\small\begin{eqnarray}
&& \frac{d^{2}G(u)}{d u^{2}}+\bigg(\frac{1}{u-1} + \frac{1}{u+\frac{3}{a^{2}\Lambda}}
+ \frac {1}{2u} \bigg)\frac{dG (u) }{d u } \nonumber \\
&& + \frac{\frac{3}{a^2\Lambda}}{4u(u+\frac{3}{a^{2} \Lambda})(u-1)}
\bigg(
\frac{(1+\frac{3}{a^2 \Lambda} )
 \frac{3}{4}\omega^{2}}{\Lambda (u+\frac{3}{a^2 \Lambda} ) }
+ \mu^{2}a^{2}u + \frac{ m^{2}
(1+\frac{3}{a^2\Lambda})}{(u-1)} + \lambda \bigg)G(u) = 0.
\end{eqnarray}
\normalsize

We gave one of the possible solutions in \cite{tolga}, for
$x= \cos^{2}\theta$ which is written as 
\begin{equation}
G(x) = x^{1/2} (x-1)^{m/2} \big(x+{\frac{{3}}{{a^2
\Lambda}}}\big)^{\frac{i\omega}{2} \sqrt{3/\Lambda}}
 H_G (a,q,\alpha,\beta,\gamma,\delta;x).
\end{equation}
Here $H_{G}$ a General Heun function ($H_G$) which is multiplied
by some  monomials to satisfy eq.(2). Thus, using the transformation  $x= \cos^{2}\theta$ in eq.(2), we get the needed singularity on the $z$ axis in \cite{tolga}.
The  solution without a singularity at the origin  for the variable $u$ will not have the square root singularity at the origin, and will have different values for the parameters for the Heun function.
 
Now we give the two solutions for our Eq. (1) explicitly. One
 of the solutions have no singularity at the origin. Since they satisfy the same equation, the solutions of both the angular and radial equations differ only in the independent variable used. If one  picks this solution, this will be  true
 also for the angular equation, Eq.(2). The other solution has a
 square root singularity at the origin.
 
 The  solution without a singularity at the origin for the
variable $u$ reads
 \begin{equation}
G(u) =  (u-1)^{\pm{m/2}} \big(u+\frac{3}{a^2
\Lambda}\big)^{\frac{{\pm}i\omega}{2} \sqrt{3/\Lambda}}
 H_G (b,q;\alpha,\beta;\gamma,\delta;u).
 \end{equation}
In  this  solution and in the solution with the singularity given below,  $m$ and ${\frac{i\omega}{2} \sqrt{3/\Lambda}}$ may take positive or negative values. In our text, we generally choose the positive value in our expressions involving these entities. 
The parameters  of the above solution are defined as
\begin{eqnarray}
b&=& -{\frac{{3}}{{a^2 \Lambda}}}, \\
-q&=&-\frac{i\omega \sqrt{3/\Lambda}}{4} + \frac{3m}{4a^2
\Lambda}+\frac{m^2}{4} (\frac{3}{a^2 \Lambda}+1)
+ \frac{3\lambda}{4a^2\Lambda}+\frac{3\omega^2}{4\Lambda}(\frac{3}{a^2 \Lambda}+1), \\
\alpha  &=& \frac{1}{2}\big( i\omega \sqrt{3/\Lambda} +m+\frac{3}{2} - \sqrt{ \frac{9}{4} - \frac{3\mu^2}{\Lambda}}\big), \\
\beta &=& \frac{1}{2}\big( i\omega \sqrt{3/\Lambda} +m+\frac{3}{2} + \sqrt{ \frac{9}{4} - \frac{3\mu^2}{\Lambda}}\big), \\
\gamma&=& \frac{1}{2}, \\ 
\delta &=& m+1, \\
u&=&\frac{-r^{2}}{a^{2}}.
\end{eqnarray}
Since the $\phi$ dependence is given
as $e^{-im\phi}$, $m$ takes only unit values.

The solution with the singularity at the origin is
\begin{equation}
 G(x) = u^{1/2} (u-1)^{m/2} \big(u+{\frac{{3}}{{a^2
\Lambda}}}\big)^{\frac{i\omega}{2} \sqrt{3/\Lambda}}
 H_G (c,Q,A,B,\Gamma,\Delta;u).
\end{equation}
 The parameters  for the equation above are defined as
\begin{eqnarray}
c&=& -{\frac{{3}}{{a^2 \Lambda}}}, \\
-Q&=&-\frac{3i\omega \sqrt{3/\Lambda}}{4} + \frac{9m}{4a^2
\Lambda}-\frac{2+6\frac{3}{4a^2\Lambda}}{4}+\frac{m^2}{4} (\frac{3}{a^2 \Lambda}+1)
+ \frac{3\lambda}{4a^2\Lambda}+\frac{3\omega^2}{4\Lambda}(\frac{3}{a^2 \Lambda}+1), \\
A  &=& \frac{1}{2}\big( i\omega \sqrt{3/\Lambda} +m+\frac{5}{2} - \sqrt{ \frac{9}{4} - \frac{3\mu^2}{\Lambda}}\big), \\
B &=& \frac{1}{2}\big( i\omega \sqrt{3/\Lambda} +m+\frac{5}{2} + \sqrt{ \frac{9}{4} - \frac{3\mu^2}{\Lambda}}\big), \\
\Gamma&=& \frac{3}{2}, \\
\Delta &=& m+1, \\
u&=&\frac{-r^{2}}{a^{2}}.
\end{eqnarray}

Both  of the wave equations  for $u$  above seem to have a singularity at $(u+\frac{3}{a^2 \Lambda })= 0$. We have to note that,
for positive
$\Lambda$, $u$ should be more than $-\frac{3}{a^2 \Lambda }$ \cite{Gibbons}.
At $u=-\frac{3}{a^2 \Lambda }$, we have the cosmological  horizon.
For negative $\Lambda$, there is no singularity at $\frac{3}{a^2 \Lambda }$, no event horizon. Eq.s (4,6) are valid for $u$ from zero to minus infinity.

In order to simplify our solutions,
we take the mass of the scalar field $\mu$ equal to zero.
When the field is massless, we get 
\begin{equation}
\alpha'  =
\frac{1}{2}( {i\omega} \sqrt{3/{\Lambda}} + {m})
\end{equation}
and
\begin{equation}
 \beta' =
\frac{1}{2}( i\omega \sqrt{3/{\Lambda}}+ m +3) .
\end{equation}
for the case without the singularity at the origin. When we have a singularity at the origin, the values of $A,B$ increase by one half.
\begin{equation}
A  =
\frac{1}{2}( i\omega \sqrt{3/{\Lambda}} + m+1)
\end{equation}
and
\begin{equation}
 B =
\frac{1}{2}(i \omega \sqrt{3/{\Lambda}}+ m +4) .
\end{equation}

We may also look for  polynomial solutions for our Heun
equation for the argument $u$. 
In
\cite{Vieira,Vieira1}, the boundary  conditions to have resonant
frequencies are given as to have the radial solution to be finite
at the horizon, in our case at the origin, and well behaved at
asymptotic infinity, which is  at the cosmological horizon for the dS or at
asymptotic infinity for the AdS cases. Vieira et al. state that to satisfy the
condition at asymptotic infinity one needs polynomial solutions. 
 The first requirement for this is to have either $A$ or $B$ equal to $-n$
\cite{Arscott1}, i.e.
, where
\begin{equation}
A  = \frac{1}{2}( i\omega \sqrt{3/{\Lambda}} +m+1)= -n.
\end{equation}
  " These frequencies are the proper modes at which a black hole freely oscillates when excited by a perturbation... Since they are damped by the emission of gravitational waves, the corresponding eigenvalues are complex \cite{Sakalli}".
$n$ is the rank of the polynomial. 
Then our solution for the variable $u$ is given by 
\begin{equation} 
u^{1/2} (u-1)^{m/2} \big(u+{\frac{3}{a^2\Lambda}\big)^{-n-\frac{1+m}{2}}}
 H_G (a,Q,A,B,\Gamma,\Delta;u).
\end{equation}
where in $A$ and $B$, we replace $i\omega \sqrt{3/{\Lambda}}$ by ${\frac{-2n-m-1}{2}}$.
The other parameter, $B$, is
given in eq.(18) above. For the generic case, i.e.  the standard form given by our eq. (3), this will be given as $\alpha+\beta+1= \gamma+\delta+\epsilon$. There is a second
necessary criterion, which is the vanishing of a determinant given
in \cite{ ciftci,karayer}.

We have to note that, when $A=-n$, we get an imaginary value for $\omega$.  Furthermore, the same $\omega$ is used in the time dependence of the solution as $e^{i\omega t}$.  We can have not a propagating but a decaying wave, since $i\omega$ does not have an imaginary part.

If we, instead, take the time dependence  of the solution as $e^{-i\omega t}$, as given in \cite{Gibbons}, we have to take negative values for $m$ and have a condition on $n$ as ${-2n-1+|m|}$ should be greater than zero, which makes the highest power of the polynomial also depend on the value of $m$. 

\section{3. Scattering Coefficient at the Origin}

There are two different cases in this section, the de Sitter and anti de Sitter cases. For the de Sitter case, we can use only a finite range of the $u$ coordinate, since we can not take $u< -3/(a^{2}\Lambda)$. Note that $u=-\frac{r^{2}}{a^{2}}$. We can use the connection formula derived in Appendix B for this case, derived, following the  method in the paper by Dekar et al.\cite{hammann}. For the anti de Sitter case, we need a connection formula relating the solution at the origin to the solution at minus infinity. As far as we know, such a connection formula for Heun functions does not exist in literature, and we could not derive one.  There is, however, a trick, used by several authors \cite{Jaffe, Philipp, Lay, Leaver, Fiziev}. We have to start by changing our variable $u$ to $v$, where this new variable is given by $v= u/(u-1)$. By doing this trick, we bring the singularity at minus infinity to $v=1$. Then, we  use the formulae derived in \cite{hammann}.

\subsection{de Sitter case}

When $\Lambda$ is positive, we can investigate if we have scattering at the origin, for waves coming
from the cosmological horizon. Since the domain of the independent variable $u$  is finite, we do not need a new transformation to limit the domain to a finite interval. Still, we investigate this case, in Appendix C, using the trick of \cite{Jaffe, Philipp, Lay, Leaver,Fiziev}, with no new results.

When $\Lambda$ is positive, we can study the scattering of the wave coming from the  cosmological
horizon at the origin, producing a reflected wave there. 
We write the wave approaching
from the point $u=-\frac{3}{a^2 \Lambda}$ to the origin. This necessitates writing
$ F(u)$ in terms of two functions with arguments $u=-\frac{3}{a^2 \Lambda}$. For the generic case, this will be
\begin{equation}
H_{G}(u) = D_{1} H_{G}(u-b) + (u-b)^{(\gamma+\delta-\alpha-\beta)} D_{2} H_{G}(u-b).
\end{equation}
Here $b,\gamma,\delta,\alpha,\beta$ are the generic parameters of General Heun equation given in our eq. (3).
We will apply this formalism to our example.

For the scalar field's mass equal zero case, one can write the  Heun solution expanded  in powers of  $u$,
 in terms of the two linearly independent Heun  solutions expanded in powers of $(u+{\frac{{3}}
 {{a^2 \Lambda}}})$:
\begin{equation}
H_{g}(u) = D_{1} H_{G}(u+{\frac{{3}}
 {{a^2 \Lambda}}}) + (u+{\frac{{3}}
 {{a^2 \Lambda}}})^{\kappa} D_{2} H_{G}(u+{\frac{{3}}
 {{a^2 \Lambda}}}),
\end{equation}
where $\kappa$ is chosen to be equal to the value to have the second solution in the Heun form \cite{Ronveaux}. Dekar et al \cite{hammann} managed
to get the formula necessary to write the $H_{G}(z)$ solution expanded in terms of $z$ in terms of two similar
functions, obtained when one expands
around $1-z$. In Appendix B, we calculate the coefficients $D_{1}$ and $D_{2}$,  applicable for expanding $H_{G} $ in terms of $H_{G}(u+{\frac{{3}}
 {{a^2 \Lambda}}})$ for expansions for the generic Heun
equation, adapted to our case.

To describe this calculation briefly, we start with our eq. (4), with  a mass zero scalar field.
For scattering for waves coming from the cosmological horizon, we take the solution, eq.(14), which is non analytic at the origin, and the second solution to this equation.
\begin{equation}
y_{i}(u) = u^{1/2} (u-1)^{{m/2}} (u+\frac{3}{a^2 \Lambda})^
{\frac{i\omega}{2} \sqrt{{3/\Lambda}}}  F_{i}(u)
\end{equation}
Here $i$ will take two values, $1,2$.

For solutions expanded in a power series in terms of $u$, we have
\begin{equation}
F_{1}(u) = H_{G}(-\frac{3}{a^2 \Lambda},q_{1}; K,J;\frac{3}{2}, m+1; u),
\end{equation}
where
\begin{equation}
 J=\frac{1}{2}((1+m+3+i\omega\sqrt{{3/\Lambda}}),
\end{equation}
\begin{equation}
 K=\frac{1}{2}(1+m+i\omega\sqrt{{3/\Lambda}}),
\end{equation}
and
\begin{equation}
q_{1}= \frac{1}{4} \big(-2 \frac{3}{a^2\Lambda}+3i\omega
\sqrt{{3/\Lambda}}+2-(\frac{3\omega^{2}}
{\Lambda}+m^{2})(1+\frac{3}{a^2 \Lambda})-(9m+\lambda))
\frac{3}{a^2\Lambda}\big)
\end{equation}
The same equation has another solution due to the properties of the Heun equation \cite{hammann1}
which has the same powers multiplying the $F_{2}$ term, and which is needed in the long calculation to obtain the connection result, given below.  This second solution reads,
 \begin{equation}
 F_{2}(u)= (u+\frac{3}{a^2 \Lambda})^
{-i\omega \sqrt{\frac{3}{\Lambda}}}
H_{G}(-{\frac{3}{a^2\Lambda}},q_{2}; L,E;;\frac{3}{2}, m+1; u),
\end{equation}
where
\begin{equation}
  E=\frac{1}{2}((1+m+3-i\omega\sqrt{{3/\Lambda}}) ,
\end{equation}
\begin{equation}
 L=\frac{1}{2}(1+m-i\omega\sqrt{{3/\Lambda}}),
\end{equation}
and
\begin{equation}
q_{2}= .\frac{1}{4} \big(-2 \frac{3}{a^2\Lambda}-3i\omega
\sqrt{{3/\Lambda}}+2-(\frac{3\omega^{2}}
{\Lambda}+m^{2})(1+\frac{3}{a^2 \Lambda})-(9m+\lambda))
\frac{3}{a^2\Lambda}\big),
\end{equation}
i.e. $q_{2}$ is the complex conjugate of  $q_{1}$. 

When expanded in terms of $u+\frac{3}{a^2\Lambda}$, we get two linearly solutions $F_{j}$ where $j$ takes two values $3,4$.
\begin{equation}
F_{3}=H_{G}(1+{\frac{3}{a^2\Lambda}}, q_{3}; K,J;
\frac{3}{2}, m+1;u+\frac{3}{a^2
\Lambda}),
\end{equation}
where
\begin{eqnarray}
&& q_{3}= \frac{1}{4} \big(6m\frac{3}{a^2\Lambda}+6mi\omega
\sqrt{{3/\Lambda}}+3mi\omega
\sqrt{{3/\Lambda}} \nonumber \\+
&& \frac{3}{a^2 \Lambda}(15i\omega
\sqrt{{3/\Lambda}}) -m^{2}-\frac{3\omega^{2}}{\Lambda}
(1+\frac{6}{a^2\Lambda})+ 2 -\lambda\frac{3}{a^2\Lambda}\big),
\end{eqnarray}
and
\begin{equation}
 F_{4}=(u+{\frac{{3}}{{a^2 \Lambda}}})^{-i\omega\sqrt{\frac{3}{\Lambda}}} 
  \times H_{G}(1+{\frac{3}{a^2\Lambda}}, q_{4}; L,E;
 \frac{3}{2}, m+1;u+\frac{3}{a^2
\Lambda}),
\end{equation}
where
\begin{eqnarray}
&& q_{4}= \frac{1}{4} \big(6m\frac{3}{a^2\Lambda}-6mi\omega
\sqrt{{3/\Lambda}}-3mi\omega
\sqrt{{3/\Lambda}}\nonumber \\ -
&& \frac{3}{a^2 \Lambda}(15i\omega
\sqrt{{3/\Lambda}}) -m^{2}-\frac{3\omega^{2}}{\Lambda}
(1+\frac{6}{a^2\Lambda})+ 2 -\lambda\frac{3}{a^2\Lambda}\big),
\end{eqnarray}
i.e. $q_{4}$ is complex conjugate of $q_{3}$.

As stated above, Dekar et al. \cite{hammann} can calculate the constants necessary  to write the General Heun function, which is a function of  the variable $z$, in terms of two solutions as functions of the variables $1-z$. 
\begin{equation}
    y_{1}(z)= C_{1} y_{3}(1-z)+ C_{2} y_{4}(1-z).
\end{equation}
Here $y_{1}(z), y_{3}(1-z),y_{4}(1-z)$ are General Heun functions with their respective parameters.
One can adapt their method   to our case.

In Appendix B, using the method of Dekar et al
\cite{hammann}, we calculated $D_{1}, D_{2}$ in terms of General Heun functions for the generic case.
Now we give the
specific values for the case we study here. Here we replaced  $D_{1}, D_{2}$ by $C_{1}, C_{2}$, to stress
that they are used not for the generic,
but for our specific case.
Then, recalling $A=K$, given in eq.s (24,33), and $B=J$, given in eq.s(25,32)
and
\begin{equation}
  E=\frac{1}{2}((1+m+3-i\omega\sqrt{{3/\Lambda}}) ,
\end{equation}
\begin{equation}
 L=\frac{1}{2}(1+m-i\omega\sqrt{{3/\Lambda}}).
\end{equation}
we get
\begin{eqnarray}
&&C_{1}\big(-{\frac{3}{a^2\Lambda}},q_{1},K,J; 
 \frac{3}{2}, m+1\big)  \nonumber \\
&=&H_{G}\big(-{\frac{3}{a^2\Lambda}},q_{1};K, J
;1+i\omega\sqrt{{3/\Lambda}},m+1;-{\frac{3}{a^2\Lambda}}\big). 
\end{eqnarray}
\begin{eqnarray}
&&C_{2}\big(-\frac{3}{a^2\Lambda},q_{1};K,J; \frac{3}{2}, m+1\big)
\nonumber \\
&=&H_{G}\big(-{\frac{3}{a^2\Lambda}},q_{1}-\frac{3}{2} i\omega
\sqrt{\frac{3}{\Lambda}};E,L ; 1-i\omega\sqrt{{3/\Lambda}},
m+1;-{\frac{3}{a^2\Lambda}}\big).
\end{eqnarray}

Using these relations, one ends up with the result
\begin{eqnarray}
&&H_{G}(-\frac{3}{a^2 \Lambda},q_{1};K,J
;\frac{3}{2}, m+1; u)=
 \nonumber\\
&& =H_{G}\big(-{\frac{3}{a^2\Lambda}},q_{1}; J, K ;
1+i\omega\sqrt{{3/\Lambda}},m+1;-{\frac{3}{a^2\Lambda}}\big) \nonumber \\
&& \times H_{G}\big(1+{\frac{3}{a^2\Lambda}}, q_{3}; K, J;
;\frac{3}{2},
 m+1;u+\frac{3}{a^2 \Lambda }\big)\nonumber \\
&& +(u+\frac{3}{a^2 \Lambda})^ {-i\omega
\sqrt{\frac{3}{\Lambda}}}H_{G}\big(-{\frac{3}{a^2\Lambda}},q_{1}-\frac{3}{2}i\omega
\sqrt{{3/\Lambda}}; E, L;  1-i\omega\sqrt{{3/\Lambda}},
,m+1;-{\frac{3}{a^2\Lambda}}\big) \nonumber \\
&&\times H_{G}\big(1+{\frac{3}{a^2\Lambda}}, q_{4};L, E;;\frac{3}{2}, m+1;u+\frac{3}{a^2 \Lambda}\big).
\end{eqnarray}
When we write this equation in the simplified form
\begin{eqnarray}
&& u^{1/2} (u-1)^{m/2} \big(u+{\frac{{3}}{{a^2
\Lambda}}}\big)^{\frac{i\omega}{2} \sqrt{{3/\Lambda}}}F_{1}= \nonumber \\
&& u^{1/2} (u-1)^{m/2} ( C_{1} \big(u+{\frac{{3}}{{a^2 \Lambda}}}\big)^{\frac{i\omega}{2} \sqrt{{3/\Lambda}}}F_{3}+
  C_{2} \big(u+{\frac{{3}}{{a^2 \Lambda}}}\big)^{-\frac{i\omega}{2} \sqrt{{3/\Lambda}}}F_{4}),
\end{eqnarray}
we see the incoming and outgoing waves. 
Then the reflection coefficient from the origin reads
\begin{equation}
R= |\frac{H_{G}(-{\frac{3}{a^2\Lambda}},q_{1}-
3/2 i\omega \sqrt{{3/\Lambda}};
E, L;
1-i\omega\sqrt{{3/\Lambda}},
,m+1;-{\frac{3}{a^2\Lambda}})))}{H_{G}(-{\frac{3}{a^2\Lambda}},q_{1};K, J;
1+i\omega\sqrt{{3/\Lambda}},m+1;-{\frac{3}{a^2\Lambda}}) }|^{2}.
\end{equation}

\subsection{Anti de Sitter case}

For the anti de Sitter case, we 
 start with our original equation, eq.(4), and  again take the mass, $\mu$,  equal to zero. To limit the domain of dependence to the interval between zero and one, we make to transformation $v= \frac{u}{u-1}$.
 This transformation will change all the terms in the original equation.

  Our eq.(4), with the transformation to $v$ reads
\small\begin{eqnarray}
&& \frac{d^{2}G(v)}{d v^{2}}+\bigg(\frac{-1}{2(v-1)} + \frac{1}{v-\frac{1}{1+\frac{a^{2}\Lambda}{3}}}
+ \frac {1}{2v} \bigg)\frac{dG (v) }{d v } \nonumber \\
&& +  \frac{3\omega^{2}}{4\Lambda} \frac {G}{{(1+\frac{a^{2}\Lambda}{3})}v ({v-\frac{1}{1+\frac{a^{2}\Lambda}{3}}})^{2}}  
-\frac {m^2 G}{4v({v-\frac{1}{1+\frac{a^{2}
\Lambda}{3}}})} + \frac{\lambda G}  {4{(1+\frac{a^{2}\Lambda}{3}})v({v-\frac{1}{1+\frac{a^{2}\Lambda}{3}}})}= 0.
\end{eqnarray}
\normalsize
The new equation is not  of the Heun form. Using a new s-homotopic tranformation by multiplying the dependent variable $H$ by $({v-\frac{1}{1+\frac{a^{2}\Lambda}{3}}})^{\kappa}$ and by choosing $\kappa$ accordingly, one can reduce the differential equation to the Heun form. This is necessary to obtain a three term recursion relation for the coefficients in our series solution.  
  
At this point, we want to demonstrate the consequences of using the analytic and non analytic solutions ot the origin. We calculate the {\it{reflection coefficients}} for these two cases. We first study a solution which is smooth at the origin, like our eq. (6). Then we study  a solution which is not analytic at the origin,like the one given in our eq. (14) and compare the results. 
 
The analytic solution will read
\begin{eqnarray}
&&G=({v-\frac{1}{1+\frac{a^{2}\Lambda}{3}}})^{\frac{i\omega}{2}\sqrt{{3/\Lambda}}} \nonumber \\ 
&& \times H_{G}(\frac{1}{1+\frac{a^2\Lambda}{3}}, q_{u};\frac{1}{2}(i\omega \sqrt{\frac{3}{\Lambda}}+m),\frac{1}{2}(i\omega \sqrt{\frac{3}{\Lambda}}-m);1/2, -1/2;v).
\end{eqnarray}
where 
\begin{equation}
q_{u}= \frac{1}{4}\bigg( -{\frac{\lambda}{1+\frac{a^2\Lambda}{3}}} +2i\omega \sqrt{\frac{3}{\Lambda}}-m^{2}-\frac{3\omega^{2}}{\Lambda}\bigg).https://tr.overleaf.com/project/5cf65feff225202797222ebf
\end{equation}
The non analytic solution at the origin will read.
 \begin{eqnarray}
&&G= ({v-\frac{1}{1+\frac{a^{2}\Lambda}{3}}})^{\frac{i\omega}{2}\sqrt{{3/\Lambda}} }\nonumber \\ 
&& \times H_{G}(\frac{1}{1+\frac{a^2\Lambda}{3}}, q_{v};\frac{1}{2}(1+i\omega \sqrt{\frac{3}{\Lambda}}+m),\frac{1}{2}(1+i\omega \sqrt{\frac{3}{\Lambda}}-m);3/2,-1/2  ; v).
\end{eqnarray}
where 
\begin{equation}
q_{v}= \frac{1}{4}\bigg( -\frac{1+\lambda}{\frac{a^2\Lambda}{3}}+ 2 +4i\omega \sqrt{\frac{3}{\Lambda}}-m^{2}-\frac{3\omega^{2}}{\Lambda}\bigg).
\end{equation}
Then we expand our solution in an infinite series, using 
 $\Sigma {a_{n} v^{n}}$,  where $v=\frac{u}{1-u}$.   
We will get three term recursion relations. As stated in \cite{Leaver, Lay}, we have to concentrate on the terms multiplying the second derivative. In the ratio test  $\frac{a_{n+1}}{a_{n}}$, which must be less than unity as $n$ goes to infinity, the important terms are the coefficients of the second derivative, since all others will be divided by $n^2$.

At the end, we find that, we have two choices. Either $\frac{a_{n+1}}{a_{n}}$ goes to unity or to $1+\frac{a^{2}\Lambda}{3}$ as $n$ goes to infinity. We choose the second choice since  $\Lambda$ is negative for this case, and $v$ is between zero and minus one. We conclude that
if the absolute value of $1+\frac{a^{2}\Lambda}{3}$ is less than unity, 
this series converges, and we can find a solution which is uniformly convergent both at zero and infinity for the variable $u$ in terms of  $v$.

Dekar et al \cite{hammann} managed
to get the formulae necessary to write the $H_{G}(z)$ solution expanded in terms of $z$ in terms of two similar
functions, obtained if one expands
around $1-z$, when the interval is finite, in their equation [A.15].
Their formula reads
\small\begin{eqnarray}
&&F(a,b;\alpha,\beta; \gamma,\delta;z) = F(a,b;\alpha,\beta; \gamma,\delta;1) F(1-a,-b-\alpha\beta;\alpha,\beta;1+\alpha+\beta-\gamma-\delta,\delta;1-z) \nonumber\\
&& +(1-z)^{\gamma+\delta-\alpha-\beta} F(a,b-a\gamma[\gamma+\delta-\alpha-\beta]; \gamma+\delta-\alpha,\gamma+\delta-\beta;\gamma,\delta;1) \times \nonumber \\
&&F(1-a,-b-\alpha\beta-[\gamma+\delta-\alpha-\beta][\gamma+\delta-a\gamma];\nonumber\\
&&\gamma+\delta-\alpha, \delta+\gamma-\beta;1+\gamma+\delta-\alpha-\beta,\delta;1-z)
\end{eqnarray}
\normalsize
The parameters in the above equation are same as the standard General Heun equation given in our eq.[3], except $q$ in that equation is equal to $-b$ here.
We use these formulae to obtain the {\it {Reflection Coefficients}} for  scattering
at the origin for both of these cases. 

For the analytic case, we get 
\small\begin{equation}
R= |\frac {\big( H_{G}(\frac{1}{1+\frac{a^2\Lambda}{3}}, q_{u}-\frac{1}{2(1+\frac{a^2\Lambda}{3})};\frac{1}{2}(-i\omega \sqrt{\frac{3}{\Lambda}}-m),\frac{1}{2}(-i\omega \sqrt{\frac{3}{\Lambda}}+m);, -1/2 ;1)\big)}
{\big(H_{G}(\frac{1}{1+\frac{a^2\Lambda}{3}}, q_{u};\frac{1}{2}(i\omega \sqrt{\frac{3}{\Lambda}}+m),\frac{1}{2}(i\omega \sqrt{\frac{3}{\Lambda}}-m);1/2,-1/2  ;1)\big)}|^{2}.
\end{equation}
\normalsize
For the non analytic case, we get 
\small\begin{equation}
R= |\frac {\big({H_{G}(\frac{1}{1+\frac{a^2\Lambda}{3}}, q_{v}-\frac{1}{2(1+\frac{a^2\Lambda}{3})},\frac{1}{2}(1-i\omega \sqrt{\frac{3}{\Lambda}}-m),\frac{1}{2}(1-i\omega \sqrt{\frac{3}{\Lambda}}+m);3/2,- 1/2 ;1)}\big)}
{\big(H_{G}(\frac{1}{1+\frac{a^2\Lambda}{3}}, q_{v};\frac{1}{2}(1+i\omega \sqrt{\frac{3}{\Lambda}}+m),(\frac{1}{2}(1+i\omega \sqrt{\frac{3}{\Lambda}}-m);3/2,-1/2  ;1)
\big)}|^{2}.
\end{equation}
\normalsize
We see that these two are distinctly different, which may be interpreted as the existence of a worm hole at the origin for the non analytic case.

Note that the formulae both for the the reflection coefficient and for the connection equation are formal expressions.  In the mathematical literature \cite{Arscott1}, technically, the solutions, we obtained, are referred as "local solutions". These solutions are analytic only within a circle which ranges from the point of expansion up to but excluding the next singularity. 
We, therefore, do not know if our solutions around $u=0$ are analytic at $u=-\frac{3}{a^2\Lambda}$ or {u} going to infinity and vice versa.
We  try to get rid of this problem by ensuring that the solution is finite at the ends by adjusting  the parameter $\Lambda$ and $a$ in our equations \cite{Jaffe,Philipp,Lay,Leaver}. The details are at these references.  We  also give another  application of this method to the dS case in Appendix C, where convergence exists for if $\frac{a^2 \Lambda}{3}$ ,is greater than unity. 

If we use polynomial solutions, then the solution will be analytic around all three singular points,  $(0,-\frac{3}{a^2\Lambda},1)$. 
Furthermore, we have to find solutions with $\omega $ which satisfy the initial criterion for a polynomial solution which is either $\alpha$ or  $\beta$ in the generic equation, eq.(3), equal to a negative integer. In our case, since there is  a half integer difference between  $\alpha$ and $\beta$, it is not possible to have both our $A$ and $B$ satisfy this criterion.  For the dS case, the reflection formula given in eq.(50) may be of value only for polynomial solutions if the independent variable does not satisfy the limitations given. 

Note that when $A=-n$ , we will get a decaying solution, which requires $\omega$ to take a negative numerical value, since we have ${\omega }$ as an exponent of one of the monomials multiplying the Heun solution. In general we may have only one of the coefficients $C_{i}$ finite, $i=1,2$ the
other infinite, given the no reflection case.

\section{4. Conclusion and Discussion}

We study the solutions  of the wave equation for a scalar particle  in the metric given in \cite{Gibbons}
for the zero mass limit of the
 Kerr (Anti) de Sitter geometry. We find two solutions, one of which has a square root singularity at the origin,
both for the radial and
 polar angle equations, which may be interpreted as the presence of a  worm hole. The other solution is
smooth at the origin.
 We find that these solutions may have polynomial solutions, since they satisfy the first criterion given in
\cite{Arscott1} for special
 values of frequency, if they also satisfy the second criterion \cite{ciftci,karayer}. We also write the reflection coefficient for waves coming from the cosmological horizon (dS) and from infinity  at the origin.
 
 In Appendix A,  we  calculate the shape of  one of the solutions  as $u$ goes to infinity, which is a finite phase, not infinite or zero. The second solution vanishes as $u$ goes to infinity.

\section{Acknowledgement}
We are grateful to Prof. Nadir Ghazanfari for informing of the the Dekar et al. paper, giving us that paper and for important discussions. We are also grateful to Prof. Ibrahim Semiz for providing us a copy of the Leaver paper. 
We thank Prof. Cemsinan Deliduman for informative discussions. We also thank Prof. Tolga Birkandan for collaboration in earlier work on this subject \cite{tolga}, and for calculating  numerical values for some of our  Heun solutions. We also thank Prof. Bekir Can L\"utf\"uoglu for technical assistance. This work  is morally supported by the Science Academy, Turkey, an NGO.

\section{Appendix A}
 
 To study the behaviour of the General Heun equation,  given in eq.(3), as $x$ goes to infinity, we transform the variable $x$
 to $y$, where $y=\frac {1}{x}$ first on the
generic Heun  equation,
\begin{equation}
\frac{{d^{2}H_G (x)}}{{d x^{2}}} +\bigg( \frac{\delta}{x-1} + \frac{\epsilon}{x-b}+ \frac{\gamma}{x}
\bigg) {\frac{{dH_G (x) }}{{d x }}}
+\bigg(\frac{\alpha \beta x-q}{x(x-1)(x-b)} \bigg) H_G(x) = 0
\end{equation}
 to see  how the standard equation changes. The above equation is transformed into
\begin{eqnarray}
&& \frac{d^{2}H_G (y)}{d y^{2}} +\bigg( \frac{\delta}{y-1} + \frac{\epsilon}{y-\frac{1}{b}}+ \frac{1-\alpha-\beta}{y}
 \bigg)
 \frac{dH_G (y)}{dy} + \frac{\alpha \beta}{y^{2}} H_G(x) \nonumber\\
&&+\bigg(\frac{-\alpha \beta y-\frac{q}{b}+ \alpha \beta (1+\frac{1}{b})}{y(y-1)(y-\frac{1}{b})} \bigg) H_G(x) = 0
\end{eqnarray}
We see that the new equation is not of the Heun form. To put it to the Heun form, we have to multiply
the solution by
 $y^{\kappa}$ where $\kappa$ is adjusted to put the equation to the Heun form. After this is done
we end up with the equation
\begin{eqnarray}
\frac{{d^{2}H (y)}}{{d y^{2}}} +\bigg( \frac{\delta}{y-1} +
\frac{\epsilon}{y- \frac{1}{b}}+
\frac{{ 2\kappa+1-\alpha-\beta}}{{y}} \bigg)
{\frac{{dH (y) }}{{dy }}}+ \nonumber \\
+\bigg(\frac{ y (\kappa(\delta+\epsilon)-\alpha \beta )-(q/b
+\frac{(\delta+b\epsilon)}{b}\kappa -\alpha \beta (1+1/b)
}{(y(y-1)(y- \frac{1}{b})} \bigg) H(y) = 0,
\end{eqnarray}
where the solution of this equation has to be multiplied by $y^{\kappa}$ to be a solution of the former equation.
Here  $\kappa$  may equal to $\alpha$ or to $\beta$, giving us
two solutions. If $\kappa= \alpha$, $\beta$ goes to
$\delta+\epsilon-\beta$. If $\kappa=\beta$, $\alpha$ goes to
$\delta+\epsilon-\alpha$. We get two solutions:
\begin{equation}
 (1/x)^{\alpha}H_{G} (1/b, q/b+ (\frac{\delta}{b}+\epsilon-\beta(1+\frac{\beta}{b}))\alpha;
 \alpha,\delta+\epsilon-\beta; \delta, \epsilon; 1/x);
\end{equation}
and
\begin{equation}
(1/x)^{\beta }H_{G} (1/b, q/b+
(\delta/{b}+\epsilon-\alpha(1+1/{b}))\beta; ,\delta+\epsilon-\alpha,\beta; \delta, \epsilon; 1/x).
\end{equation}
If we want the behavior at infinity, we see that the equation behaves
as $ x^{-\alpha} $ or $ x^{-\beta }$ plus terms with higher powers
of $ 1/x $, since for $\frac{1}{x}=y=0$ the Heun solution $H_{y=0}$  is unity.

All through these calculations we used the Heun equation  constraint $ \alpha+\beta+1= \gamma
 +\delta+\epsilon $ on the  parameters of the standard form in terms of the variable $x$. Actually, this is constraint does not really constrain anything and is a result of using not a single parameter, but the product of two parameters,$\alpha \times \beta$, in the standard form of the General Heun equation given in our eq.(3), in the term multiplying $x$.

To apply this result to our  solution when $u$ goes to minus infinity, we first have to put our eq.(4) to the Heun form. This gives us two solutions,  eq.s(6,14). Then we make  the transformation $u= \frac{1}{s}$ on this transformed differential equation, eq.(14), and put the resulting equation  to the Heun form to
 obtain the solution
\begin{equation}
G(s) = u^{1/2}(u-1)^{{m/2}} \big(u+\frac{3}{a^2 
\Lambda}\big)^{\frac{i\omega}{2} \sqrt{3/\Lambda}}
(s)^{{{\frac{i\omega}{2} \sqrt{3/ \Lambda}}}} (s)^{{m/2}} s^{1/2} ) H_{G}
\end{equation}
\normalsize
where $H_{G}$ is given as
\begin{equation}
H_{G}( -{\frac{a^{2} \Lambda}{3}}, \frac{1}{4} ( \lambda +i\omega \sqrt{\frac{\Lambda}{3}}; \frac{1}{2} ( m+i\omega \sqrt{\frac{\Lambda}{3}}),
 \frac{1}{2} ( m+i\omega \sqrt{\frac{\Lambda}{3}}+1); -\frac{1}{2}, m+1 ; s).
\end{equation}
When 
 $s$ equals zero, we see that there is only a constant term, a phase, equal to
$e^{i\pi\frac{m}{4}}( \frac{a^2\Lambda}{3})^{-i\omega
\sqrt{3/{\Lambda}}}$. The monomials multiplying $H_{G}$ cancel out
and $H_{G}$ equals unity for $s=0$. 
Since the Heun function goes to unity at $s=0$, this validates our result given above. We see that the solution is smooth as $u$ goes to infinity.
Here we used only the solution with $\alpha$.
 There is another solution which goes to zero for $s=0$, which corresponds to the solution given in eq.(63). The solution  with $\beta$ has a square root  singularity at $s=0$.

 In the main text, we use the method of \cite{Jaffe, Lay, Philipp, Leaver}  in addition to the Dekar et al \cite{hammann} to study scattering of a wave coming from infinity.

\section{Appendix B}

Dekar et al  could write the  General Heun solution for the variable $z$  in terms of two independent solutions for the variable $1-z$ \cite{hammann} as
\begin{equation}
    y_{1}(z)= C_{1} y_{3}(1-z)+ C_{2} y_{4}(1-z)
\end{equation}
where
\begin{equation}
y_{1}(z)= H_{G}( a,q;\alpha,\beta;\gamma, \delta; z)
\end{equation},
 \begin{equation}
 y_{3}(1-z)= H_{G}(1-a, -q - \alpha \beta ; \alpha,\beta; 1 \alpha+\beta-\gamma- \delta;\delta ; 1-z),
\end{equation}
\begin{equation}
y_{4}(1-z)= (1-z)^{(-\alpha-\beta+\gamma+ \delta)}
H_{G}(1-a,Q^{+};
\gamma+\delta-\alpha, \gamma+\delta-\beta; 1+\gamma+\delta -\alpha-\beta, \delta; 1-z).
\end{equation}
Here
\begin{equation}
Q^{+}= -q -\alpha \beta+(\alpha+\beta-\gamma- \delta)(\gamma+\delta-a\gamma).
\end{equation}
Here $ C_{1}$ and $C_{2}$ are constants that depend on the parameters of the $ H_{G}$ function and $y_{1}$ satisfies
the General Heun equation,
\begin{eqnarray}
&&\frac{{d^{2}H_G (z)}}{{d z^{2}}} +\bigg( \frac{\delta}{z-1} + \frac{\epsilon}{z-a}+ \frac{{\gamma}}{{z}} \bigg)
{\frac{{dH_G (z) }}{{d z }}}
+\bigg(\frac{{\alpha \beta z-q}}{{z(z-1)(z-a)}} \bigg) H_G(z) = 0.
\end{eqnarray}
By using a clever trick they can evaluate the constants $ C_{1}$ and
$ C_{2}$ as
\begin{equation}
C_{1} (a,q; \alpha,\beta;\gamma,\delta) = H_{G}( a,q;\alpha,\beta;\gamma, \delta; 1)
\end{equation}
and
\begin{equation}
C_{2} (a,q; \alpha,\beta,\gamma,\delta) =  H_{G} (a,q+a\gamma (\alpha+\beta-\gamma-\delta),\gamma+\delta-\alpha,
\gamma+\delta-\beta;\gamma,\delta;1).
\end{equation}
One can use the same method to write
\begin{equation}
y_{1}= H_{G}( a,q;\alpha,\beta;\gamma, \delta;z)
\end{equation}
in terms of two independent solutions expanded around $z-a$.

\begin{equation}
y_{1}= H_{G}( a,q;\alpha,\beta;\gamma, \delta; z) =D_{1} y_{5}(z-a)+ D_{2} y_{6}(z-a).
\end{equation}
where $ D_{1}, D_{2}$ are functions of the  parameters of the   $H_{G}$ function on the left hand side of
the above equation.
Here
\begin{equation}
 y_{5}= H_{G}(1-a, q - a\alpha \beta ; \alpha,\beta; 1+ \alpha+\beta-\gamma- \delta;\delta ; z-a),
\end{equation}
and
\begin{equation}
 y_{6}= (z-a)^{(-\alpha-\beta+\gamma+ \delta)}\nonumber 
 H_{G}(1-a, Q^{*};
\gamma+\delta-\alpha, \gamma+\delta-\beta; 1+\gamma+\delta -\alpha-\beta, \delta;z-a),
\end{equation}
where 
\begin{equation}
 Q^{*} = q-a\alpha \beta+(\alpha+\beta-\gamma- \delta)(a(\gamma+\delta)-\gamma).
 \end{equation}
 Then we write the equality
 \begin{eqnarray}
&&y_{1}= H_{G}( a,q;\alpha,\beta;\gamma, \delta; z)= \nonumber \\
&& (z-a)^{(-\alpha-\beta+\gamma+ \delta)}
 H_{G}(1-a, q-a\alpha \beta (\alpha+\beta-\gamma- \delta)(a(\gamma+\delta)-\gamma);\alpha, \beta;\gamma,\delta; z).
 \end{eqnarray}
 An equality similar to this is given in \cite{hammann} for expansion around $1-z$ with the proper citation.
 Then we use the expansion given in our eq.(75) where we write $D_{i}$ with proper indices.
\begin{eqnarray} 
&& H_{G}( a,q;\alpha,\beta;\gamma, \delta;z)=\nonumber \\
&&D_{1} (a,q;\alpha,\beta;\gamma,\delta)  H_{G}(1-a, q - a\alpha \beta ; \alpha,\beta; 1+ \alpha+\beta-\gamma- \delta;\delta ; z-a) \nonumber \\
&&+ D_{2} (a,q;\alpha,\beta;\gamma,\delta) 
(z-a)^{(-\alpha-\beta+\gamma+ \delta)}\nonumber \\
 &&\times H_{G}(1-a, Q^{*};
\gamma+\delta-\alpha, \gamma+\delta-\beta; 1+\gamma+\delta -\alpha-\beta, \delta;z-a).
\end{eqnarray}
Then we note that if $\gamma+\delta$ is greater than $\alpha+\beta$, we get
\begin{equation}
 H_{G}( a,q;\alpha,\beta;\gamma, \delta;a)= D_{1} (a,q;\alpha,\beta;\gamma,\delta)    
\end{equation},
since when $z=a$, $H_{G}(1-a, q - a\alpha \beta ; \alpha,\beta; 1+ \alpha+\beta-\gamma- \delta;\delta ; 0)=1$.
We, then,  let $\alpha$ go to $\gamma+\delta-\alpha$ and $\beta$ go to $\gamma+\delta-\beta$ in eqs. (79,80). We get
\begin{eqnarray}
 &&H_{G}( a,q;\gamma+\delta-\alpha,\gamma+\delta-\beta;\gamma, \delta; z)= \nonumber \\
&& (z-a)^{(+\alpha+\beta-\gamma- \delta)}
 H_{G}(a, q+\gamma(\alpha+\beta-\gamma- \delta);\alpha, \beta;\gamma,\delta; z).
 \end{eqnarray}   
 and
\begin{eqnarray} 
&&
H_{G}( a,q;\gamma+\delta-\alpha,\gamma+\delta-\beta;\gamma, \delta; z)=\nonumber \\
&& (z-a)^{(\alpha+\beta-\gamma-\delta)}
 H_{G}(a, q+\gamma(\alpha+\beta-\gamma- \delta);\alpha, \beta;\gamma,\delta; z)=\nonumber \\
&&(z-a)^{(+\alpha+\beta-\gamma- \delta)} \nonumber\\
&&\times \Big( 
D_{1} (a,q+\gamma(\alpha+\beta-\gamma- \delta);\alpha, \beta;\gamma,\delta) \nonumber \\ &&H_{G}( 1-a,q+\gamma(\alpha+\beta-\gamma- \delta)-a\alpha \beta;\alpha,\beta;1+\alpha4\beta-\gamma-\delta, \delta;z-a)   \nonumber \\
&&+ D_{2} (a,q+\gamma(\alpha+\beta-\gamma- \delta);\alpha, \beta;\gamma,\delta) 
(z-a)^{(-\alpha-\beta+\gamma+ \delta)}\nonumber \\
 &&\times H_{G}(1-a,Q^{**};
\gamma+\delta-\alpha, \gamma+\delta-\beta; 1+\gamma+\delta -\alpha-\beta, \delta;z-a)\Big).
\end{eqnarray}
where 
\begin{equation}
Q^{**} =q-a(\gamma+\delta-\alpha)(\gamma+\delta-\beta).  
\end{equation}
When we compare eqs.(80) and (83), we see that  \begin{equation}
D_{2}(a,q+\gamma(\alpha+\beta-\gamma- \delta);\alpha, \beta;\gamma,\delta)= 
D_{1}(a,q;\gamma+\delta-\alpha,\gamma+\delta-\beta;\gamma,\delta).
\end{equation}
Using eq.(81) and properly translation constants we get 
\begin{equation}
 D_{1}( a,q;\alpha,\beta;\gamma, \delta) = H_{G}( a,q;\alpha,\beta;\gamma, \delta; a)
 \end{equation}

\begin{equation}
  D_{2}( a,q;\alpha,\beta;\gamma, \delta) =  
  H_{G}( a,q-\gamma(\alpha+\beta-\gamma- \delta);\gamma+ \delta-\alpha,;\gamma+ \delta-\beta; a).
 \end{equation}
We use this result, with appropriate changes in the parameters to fit to our example, in Section 3.

\section{Appendix C}

Here we will find another solution for the differential equation given in our equation (4), whose  wormhole solution was given in equations ( 14-21). We were not certain that for our eq. (4), a valid solution at one singularity may not diverge at  another singularity.

The problem of finding a solution to a differential  equation which is not singular at two  consecutive singular  points was solved by mathematicians long ago. One can find the method and references to original work in \cite{Lay}. 
We had used this method to get a converging solution for the AdS case in Section 3 in the main text. Here we do the same thing for the dS case. This calculation dublicates the results given for the dS case, in the main text, since the interval for the independent variable for this case is finite, and only gives a limit on the values of $a^{2}$
and $\Lambda$. To find  a solution which is uniformly convergent both at the origin and at the cosmic horizon, $u=-\frac{3}{a^{2}\Lambda}$, we transform to a new variable $z={\frac{u+{\frac{3}{a^{2}\Lambda}}} {1-u}}$. This transformation brings $u=0$ to $z={\frac{3}{a^{2}\Lambda}}$, and $u=-{\frac{3}{a^{2}\Lambda}}$ to $z=0$.  We go through  similar steps, as in the AdS case, transform the new differential equation to the Heun form by a  s-homotopic transformation by multiplying the dependent variable by $(1+z)^{\nu}$. Then we expand our solution as a power series expansion around $z=0$. and find a series solution that converges uniformly if $\frac{a^{2}\Lambda}{3}$ is greater than unity. Thus, one can obtain a convergent solution at the both singular points of our respective domains of interest. 
The solution will read
\begin{eqnarray}
&& M= u^{1/2} (u-1)^{-1/2(i\omega \sqrt{3/{\Lambda}}}) (u+\frac{3}{a^2 \Lambda})^{\frac{i\omega}{2} \sqrt{{3/\Lambda}}} \nonumber \\
&& \times H_{G}(\frac{3}{a^2 \Lambda}),Q_{z}; 1/2(m+i\omega \sqrt{{3}{\Lambda}}),1/2(-m+i\omega \sqrt{\frac {3}{\Lambda}} ;1+i\omega \sqrt{\frac{3}{\Lambda}},-3/2;\frac{(u+\frac{3}{a^2 \Lambda})}{1-u}),
\end{eqnarray}
where 
\begin{equation}
4Q_{z} = m^{2} + \frac{3\omega^{2}}{\Lambda} +(1+ \frac{3}{a^2 \Lambda})(2-3i\omega \frac{3}{\Lambda})-\frac{3}{a^2 \Lambda}\lambda. 
\end{equation}

We do not the study the analytic solution for the de Sitter case, thinking that the reflection coefficient will  be different from the non analytic case anyway, since the solutions will be different.

Note also  that when $u=1$, the remaining singular point is not in our domain of interest, since $u$ was defined as  $ u= - r^2/a^2 $. When $u=1$, our physically interesting variable $r$ takes complex values.

\end{document}